\documentclass[prl,twocolumn,amsmath,amssymb]{revtex4}
\newcommand{\be}{\begin{equation}}
\newcommand{\ee}{\end{equation}}
\newcommand{\ba}{\begin{eqnarray}}
\newcommand{\ea}{\end{eqnarray}}
\newcommand{\nn}{\nonumber}

\usepackage{graphicx}
\usepackage{bm}

\begin{document}

\title{Quasiparticle Decoherence in $d$-wave Superconducting Qubits}
\author{M.~H.~S.~Amin and A.~Yu.~Smirnov}
\affiliation{%
D-Wave Systems Inc., 320-1985 W. Broadway, Vancouver, B.C., V6J
4Y3 Canada}


\begin{abstract}
It is usually argued that the presence of gapless quasiparticle
excitations at the nodes of the $d$-wave superconducting gap
should strongly decohere the quantum states of a $d$-wave qubit,
making quantum effects practically unobservable. Using a
self-consistent linear response non-equilibrium quasiclassical
formalism, we show that this is not necessarily true. We find
quasiparticle conductance of a $d$-wave grain boundary junction
to be strongly phase dependent. Midgap states as well as nodal
quasiparticles contribute to the conductance and therefore
decoherence. Quantum behavior is estimated to be detectable in a
qubit containing a $d$-wave junction with appropriate parameters.
\end{abstract}
\vspace{5mm}


\maketitle

Among numerous qubit implementations, superconducting ones enjoy
long decoherence times because of their gapped electronic
excitation spectrum. This fact has recently been confirmed by
several striking experiments \cite{expt}. The key constituents in
all of those are Josephson tunnel junctions. One advantage of the
tunnel junctions is the exponential dependence of their
quasiparticle resistance $R$ on temperature $T$ (i.e. $R \sim
e^{\Delta/T}$ where $\Delta$ is the superconducting gap. Herein
$k_B {=} \hbar {=} 1$). The electronic decoherence is therefore
exponentially suppressed at low $T$. Similar behavior also exists
in superconducting point contacts, with the energy of Andreev
levels $\epsilon_0(\phi)=\Delta \cos \phi/2$, replacing $\Delta$
in the exponent \cite{LY,new}. Deviation from the exponential
dependence is expected at low temperatures \cite{new}.

Despite their naturally degenerate ground states
\cite{ilichev01}, desirable for quantum computation, $d$-wave
qubits \cite{dwqubit,SPQ} are controversial because their
quasiparticle spectrum is gapless at the nodes of the order
parameter. Moreover, experimentally, the normal resistance
extracted from I-V characteristics of $d$-wave grain boundary
junctions is found to be very small \cite{HM,RC} (with $RC \sim
1$ps \cite{RC}); much smaller than required to observe quantum
effects. However, the resistance is measured in the running state
of the junctions. The Doppler shift due to large superconducting
current in such a state will populate the nodes, enhancing the
conductance $G$ ($\equiv 1/R$). Moreover, time variation of the
phase difference across the junction would effectively {\em phase
average} $G$. As we shall see, midgap states (MGS) can
significantly contribute to such an averaged $G$, except at very
low $T$.

The first step is therefore to calculate $G$ for a $d$-wave grain
boundary junction. Most existing methods can study ac properties
of a Josephson junction biased with a constant voltage (see
e.g.~\cite{hurd,zaikin}). This however implies a constant
variation of phase difference $\phi$, which is not the case in
qubits. In Ref.~\onlinecite{new}, a self-consistent
non-equilibrium quasiclassical technique was developed to
calculate linear response of a Josephson junction to an ac
voltage with arbitrary frequency. The method was successfully
applied to the case of a superconducting point contact. In high
$T_c$ superconductors, the relatively large $T_c/E_F$ ($E_F$ is
the Fermi energy), makes the quasiclassical approximation only
marginally applicable. It nevertheless has proven successful in
calculating equilibrium properties of $d$-wave superconductors
\cite{dwave,amin,MS}. Here, we employ the theory of
Ref.~\onlinecite{new} to calculate $G$, and therefore
decoherence, in a $d$-wave grain boundary junction.

Let us now briefly describe the technique. More details are
provided in Ref.~\onlinecite{new}. We calculate quasiclassical
retarded, advanced, and Keldysh Green's functions
\cite{new,eschrig}
\begin{eqnarray}
\widehat{g}^{R,A}=\left(
\begin{array}{cc}
g^{R,A} & f^{R,A} \\
f^{\dagger R,A} & -g^{\dagger R,A}
\end{array}
\right) , \quad \widehat{g}^K=\left(
\begin{array}{cc}
g^K & f^K \\
-f^{\dagger K} & g^{\dagger K}
\end{array}
\right), \nonumber
\end{eqnarray}
which are functions of the Fermi velocity ${\bf v}_{F}$,
quasiparticle energy $\epsilon$, position ${\bf r}$, and time
$t$. Here $f^{\dagger R} ({\bf v}_F,\epsilon;{\bf r},t) \equiv
f^R(-{\bf v}_F,-\epsilon;{\bf r},t)^*$, etc. In equilibrium, the
retarded and advanced Green's functions can be written in terms
of Riccati amplitudes \cite{MS} $a_0^\alpha$ and $b_0^\alpha$ in
a way very similar to the conventional method for the Matsubara
Green's functions \cite{MS,amin}:
\begin{equation}
g_0^\alpha= s^\alpha {1 - a_0^\alpha b_0^\alpha \over 1 +
a_0^\alpha b_0^\alpha} \ , \qquad f_0^\alpha= s^\alpha {2
a_0^\alpha \over 1 + a_0^\alpha b_0^\alpha }, \label{gf0RA}
\end{equation}
where $\alpha = R,A$ for retarded and advanced functions
respectively, and $s^\alpha = +$ ($-$) for $\alpha=R$ ($A$). The
subscript ``0'' denotes equilibrium quantities. The amplitudes
satisfy Riccati-type equations
\begin{eqnarray}
{\bf v}_{F}\cdot \nabla a_0^\alpha &=& 2i \epsilon^\alpha
a_0^\alpha -
(a_0^\alpha)^{2} \Delta^*_0 +  \Delta_0, \nonumber \\
-{\bf v}_{F}\cdot \nabla b_0^\alpha &=& 2i \epsilon^\alpha
b_0^\alpha - (b_0^\alpha)^{2}\Delta_0 + \Delta^*_0 , \label{rcti}
\end{eqnarray}
where $\epsilon^\alpha = \epsilon + i s^\alpha \eta$, with
$\epsilon$ and $\eta$ being the real and imaginary parts of the
quasiparticle energy respectively. In $d$-wave superconductors,
$\eta$ results from both inelastic and impurity scattering
processes \cite{eta}. The boundary conditions are the bulk
solutions of (\ref{rcti}): $a_0^\alpha = \Delta_0
/(-i\epsilon^\alpha + s^\alpha \Omega^\alpha)$ and $b_0^\alpha =
\Delta^*_0 / (-i\epsilon^\alpha + s^\alpha \Omega^\alpha)$, where
$\Omega^\alpha {=} \sqrt{|\Delta_0|^2 - (\epsilon^{\alpha})^2 }$.
To calculate $a_0^R$ and $b_0^A$ ($b_0^R$ and $a_0^A$), we
integrate (\ref{rcti}) in the direction of ${\bf v}_F$ ($-{\bf
v}_F$), starting from the boundary conditions at $x{=}-\infty$
($+\infty$).

Non-equilibrium behavior emerges when an ac voltage $V(t)=V_0 \cos
\omega t$ is applied across the junction. Choosing a gauge in
which the vector potential ${\bf A}=0$, we consider a scalar
potential $\Phi= \pm (V_0/2) \cos \omega t$, with the $+ \ (-)$
sign on the left (right) side of the junction. To ensure
applicability of the linear response formalism, it is necessary
that $eV_0 \ll \Delta,\omega$ \cite{note}.

We define the linear response Green's functions $\delta
\widehat{g}^\alpha = \widehat{g}^\alpha - \widehat{g}_0^\alpha$,
related to the linear response amplitudes $\delta a^\alpha =
a^\alpha - a_0^\alpha$ and $\delta b^\alpha = b^\alpha -
b_0^\alpha$ through ($\alpha = R,A$)
\begin{eqnarray}
\delta g^\alpha &=& -2 s^\alpha {\delta a^\alpha b_{0-}^\alpha +
\delta b^\alpha a_{0+}^\alpha \over (1 + a_{0+}^\alpha
b_{0+}^\alpha)(1 + a_{0-}^\alpha b_{0-}^\alpha)}, \\
\delta f^\alpha &=& 2 s^\alpha {\delta a^\alpha  - \delta b^\alpha
a_{0+}^\alpha a_{0-}^\alpha \over (1 + a_{0+}^\alpha
b_{0+}^\alpha)(1 + a_{0-}^\alpha b_{0-}^\alpha)},
\end{eqnarray}
where $a^\alpha_{0\pm} \equiv a^\alpha_0 (\epsilon \pm \omega/2)$,
etc. We also introduce the anomalous Green's function $\delta
\widehat{g}^X$ (with the same matrix form as $\widehat{g}^R$) by
$\delta \widehat{g}^K = \delta \widehat{g}^X ({\cal F}_+ - {\cal
F}_-) + \delta \widehat{g}^R {\cal F}_- - \delta \widehat{g}^A
{\cal F}_+$, where ${\cal F}_\pm \equiv \tanh[(\epsilon \pm
\omega/2)/
 2 T]$. Correspondingly, we introduce anomalous functions
$\delta a^X$ and $\delta b^X$ which are related to the Green's
functions through
\begin{eqnarray}
\delta g^X &=& 2 {\delta a^X - \delta b^X a_{0+}^{R} b_{0-}^{A}
\over (1 + a_{0+}^{R} b_{0+}^{R})(1 + a_{0-}^{A} b_{0-}^{A})}, \\
\delta f^X &=& 2 {\delta a^X a_{0-}^{A} + \delta b^X a_{0+}^{R}
\over (1 + a_{0+}^{R} b_{0+}^{R})(1 + a_{0-}^{A} b_{0-}^{A})}.
\end{eqnarray}
The differential equations describing $\delta a^\alpha$ are \
${\bf v}_F \cdot \nabla \delta a^\alpha = A^\alpha \delta
a^\alpha + B^\alpha$, with bulk boundary conditions $\delta
a^\alpha = -B^\alpha / A^\alpha$, where
\begin{eqnarray}
&& \hspace{-4mm} A^\alpha {=} \left\{ \begin{array}{cc}
2i \epsilon - (a_{0+}^\alpha + a_{0-}^\alpha) \Delta_0^\dagger, &  \alpha=R,A \\
i\omega - a_{0+}^R  \Delta_0^\dagger + b_{0-}^A \Delta_0, &
 \alpha= X
\end{array} \right. \nn \\
&& \hspace{-4mm} B^\alpha {=} \left\{ \begin{array}{cc} \delta
\Delta -a_{0+}^\alpha a_{0-}^\alpha \delta
\Delta^\dagger -ie\Phi (a_{0+}^\alpha - a_{0-}^\alpha), &  \alpha=R,A \\
a_{0+}^R \delta \Delta^\dagger +  b_{0-}^A \delta \Delta - ie\Phi
(1 + a_{0+}^R  b_{0-}^A). &  \alpha= X
\end{array} \right. \nn
\end{eqnarray}
The corresponding equations for $\delta b^\alpha$ can be obtained
by applying a $\dagger$-operation to both sides of the above
equations. Integrations for $\delta a^R$, $\delta b^A$, and
$\delta a^X$ ($\delta b^R$, $\delta a^A$, and $\delta b^X$) are
taken in the direction of ${\bf v}_F$ ($-{\bf v}_F$), along the
quasiclassical trajectories.

The equilibrium order parameter, $\Delta_0$, is calculated self
consistently using numerical iteration, while the linear response
$\delta \Delta \equiv \Delta - \Delta_0$ is given by \cite{new}
\begin{equation}
\delta \Delta (\omega) = -2e (\Delta_0/ \omega) \Phi (\omega).
\label{dd}
\end{equation}
$\delta \Delta$ satisfies the self-consistency relation up to an
error of O$(j_c/j_{\rm c,bulk})$, where $j_c$ ($j_{\rm c,bulk}$)
is the Josephson (bulk) critical current density. In $d$-wave
grain boundary junctions, $j_c/j_{\rm c,bulk}$ is smallest for a
$0^\circ{-}45^\circ$ junction. Application of (\ref{dd}), instead
of common self-consistent iterative methods \cite{eschrig}, is a
clear advantage of the present technique.

\begin{figure}
\includegraphics[width=8.5cm]{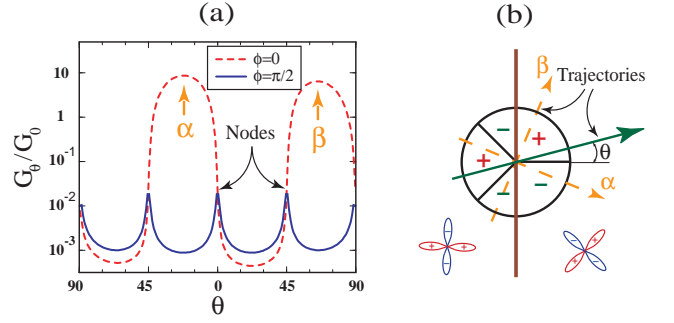}
\caption{\label{fig1} (a) Angle resolved conductance, normalized
to $G_0 = e^2 N_F v_F S$, for $\phi=0$ (dashed line) and
$\phi=\pi/2$ (solid line) at $T=0.01T_c$. (b) Quasiparticle
trajectories crossing the grain boundary. Trajectories along
$\alpha$ and $\beta$ directions see sign change of the order
parameter. Other parameters are: $\omega = 0.02T_c$ and $\eta =
0.05T_c$. }
\end{figure}

We consider an asymmetric $0^\circ{-}45^\circ$ misoriented
$d$-wave grain boundary junction, with perfect transparency and no
roughness. To clearly see the effect of the nodes and bound
states, it is useful to define an angle resolved conductance
$G_\theta = {\rm Re}[\delta j_\theta/V_0]S$, where $S$ is the the
area of the junction and
\begin{equation}
\delta  j_\theta = {e v_F N_F \over 4}
\int_{-\epsilon_c}^{\epsilon_c} d\epsilon   {\rm Tr}
\left[\widehat{\tau}_3 \delta \widehat{g}^K \right]_{{\bf v}_F =
v_F \hat{\bf n}} \label{dj}
\end{equation}
%
%
%
is the contribution to the ac current density from a trajectory
in $\theta$ direction (cf. Fig.~\ref{fig1}b). Here
$\widehat{\tau}_3$ is the third Pauli matrix, $\hat{\bf n}$ is a
unit vector in the direction of the trajectory ($\hat{\bf n} \cdot
\hat{\bf x} =\cos \theta$), $N_F$ is the density of states at the
Fermi surface, and $\epsilon_c$ is an energy cutoff, which in our
numerical calculations is taken to be 10$T_c$. (The results
however are insensitive to the exact value of $\epsilon_c$.) The
total conductance is given by $G= \int (d\theta/2\pi) G_\theta
\cos \theta$.

Figure~\ref{fig1}a shows $G_\theta$ for two values of phase
difference $\phi$ across the junction. At $\phi=0$, $G$ is
maximum at $\theta = -22.5^\circ$ and $67.5^\circ$ (paths
$\alpha$ and $\beta$ in Fig.~\ref{fig1}b). These are the
directions along which the quasiparticles see different signs of
the $d$-wave order parameter on either side of the boundary. Zero
energy Andreev bound states formed in these directions are
responsible for the large $G$ \cite{new}. Notice the asymmetry of
$G_\theta$ with respect to $\theta=0$ at $\phi=0$. This asymmetry
leads to an ac current parallel to the boundary \cite{hurd}. The
current vanishes at $\phi=\pi/2$ and is not dissipative (it is
perpendicular to the electric field), therefore not central to
our discussion.

A nonzero phase difference splits the bound states, moving them
away from zero energy. Maximum splitting occurs at $\phi=\pi/2$.
Nodal directions dominate the quasiparticle current at this phase
difference (cf. Fig.~\ref{fig1}a). Notice orders of magnitude
difference in the conductance maxima between the two phase
differences.

\begin{figure}[t]
\includegraphics[width=8.7cm]{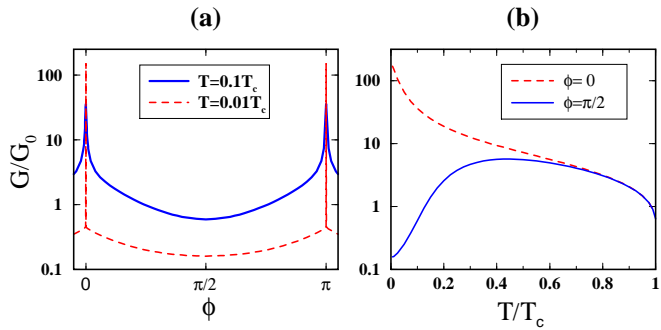}
\caption{\label{fig2} $\phi$ (a) and $T$ (b) dependence of $G$ for
the same set of parameters as in Fig.~\ref{fig1}. }
\end{figure}

\begin{figure}[t]
\includegraphics[width=8.7cm]{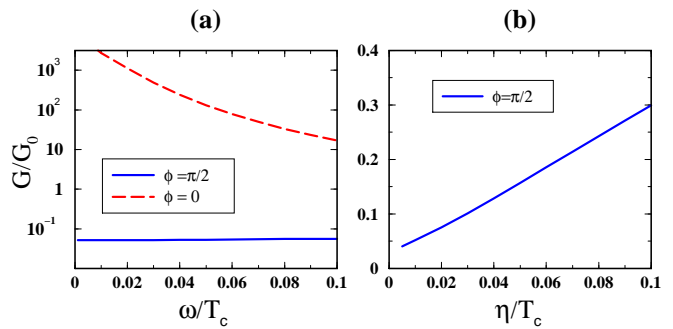}
\caption{\label{fig3} Dependence of $G$ on $\omega$ (a) and $\eta$
(b) at $T=10^{-4}T_c$. $\eta$ in (a) and $\omega$ in (b) are
equal to $0.01T_c$. }
\end{figure}

Figure~\ref{fig2}a displays $\phi$-dependence of the total $G$ at
two temperatures. Sharp peaks at $\phi = 0$ and $\pi$, 2--3 orders
of magnitude larger than the minimum (at $\phi = \pi/2$), result
from the MGS. The conductance spikes are much narrower at
$T=0.01T_c$ than at $0.1T_c$ and become extremely sharp at even
lower $T$. As a result, the influence of the MGS is small at low
$T$ (see below). Figure~\ref{fig2}b shows the $T$-dependence of
$G$ at $\phi=0$ and $\pi/2$. While the two curves coincide at high
$T$, they diverge as $T$ is lowered. The conductance at
$\phi=\pi/2$ is suppressed at low $T$. This is because the
gapless quasiparticles on one side of the junction see a gapped
order parameter on the other side \cite{bruder}. As $T
\rightarrow 0$, however, it starts to saturate to a value almost
proportional to $\eta$ (see Fig.~\ref{fig3}b). Similar behavior
was also found for the case of superconducting point contacts
\cite{new}.

To reduce decoherence, the qubit should be operated at low $T$.
Figure~\ref{fig3} shows the $\omega$ and $\eta$ dependence of $G$
at $T=10^{-4}T_c$. While strong $\omega$-dependence of $G$ exists
at $\phi = 0$, it is almost frequency independent when $\phi =
\pi/2$ (Fig.~\ref{fig3}a). The $\eta$-dependence of $G$ at
$\phi=\pi/2$, on the other hand, is close to linear
(Fig.~\ref{fig3}b). The positive slope can be attributed to the
broadening of the MGS with $\eta$, which enhances their overlap
near zero energy, increasing their contribution to $G$.

We now try to calculate the decoherence time in a $d$-wave qubit,
assuming a double well potential $U(\phi)$ with minima at $\pm
\phi_0 \approx \pm \pi/2$. (A practical example of such a system
is given in Ref.~\onlinecite{SPQ}.) We first write down a
Hamiltonian which, in
classical regime, reproduces the above calculated $G$. 
At low frequencies, $G$ shows slow $\omega$-dependence for almost
all $\phi$ (except for $\phi=0,\pi$, which do not contribute to
decoherence---see below). One can therefore assume our system to
be coupled to an Ohmic heat bath \cite{leggett}. To get a
$\phi$-dependent $G$, however, it is necessary to consider {\em
nonlinear} coupling to the bath \cite{caldeira,smirnov} [see
Eq.~(\ref{F}) below]. We therefore write the Hamiltonian as
\begin{equation}
 H = {Q^2 \over 2C} + U(\phi) - F(\phi) X,  \label{H}
\end{equation}
where $Q=2eP_\phi$ ($P_\phi$ momentum conjugate to $\phi$) is the
charge stored on the junction, and $X$ is a heat bath operator.
To find the relation between the nonlinear coupling function $F$
and $G$, we first study the classical behavior of the above
Hamiltonian. The equation of motion is \ba
 (C / 4e^2) \ddot{\phi} = - \partial_\phi U(\phi)
 + \partial_\phi F(\phi) X, \label{ddphi0}
\ea
Assuming small back-action of the system on the heat bath, one can
use linear response theory \cite{smirnov}  to write
\ba
 X(t) = X^{(0)}(t) + \int dt' D(t-t') F[\phi(t')], \label{LRT}
\ea
where $X^{(0)}$ is the unperturbed bath operator, $D(t-t') =
\langle i[X^{(0)}(t),X^{(0)}(t')]_- \rangle \theta(t-t')$ is the
bath's retarded Green's function, and $\langle ... \rangle$
denotes equilibrium statistical averaging. For an Ohmic heat bath
$D(t)=-\alpha \delta'(t)$ [in Fourier space $D(\omega)=i\alpha
\omega$], with $\alpha$ being the dissipation coefficient and
$\delta'(t)$ defined by $\int dt' \delta'(t-t') z(t') =
\dot{z}(t)$.

Eq.~(\ref{ddphi0}) therefore leads to a stochastic equation that
resembles the classical Langevin equation
\ba
 (C / 4e^2) \ddot{\phi} + \alpha \left( \partial_\phi F
 \right)^2 \dot{\phi} + \partial_\phi U
 = \xi_\phi, \label{langevin}
\ea
where $\xi_\phi = (\partial_\phi F) X^{(0)}(t)$ is the
fluctuation force \cite{xi}. The second term in (\ref{langevin})
gives dissipation and therefore corresponds to the
$\phi$-dependent $G$ \cite{caldeira}: $
 G(\phi) = 4e^2\alpha \left( {\partial_\phi F} \right)^2.
$ As a result
\ba
 F(\phi) =  {1 \over 2e \sqrt{\alpha}} \int_0^\phi d\phi'
 \sqrt{G(\phi')}. \label{F}
\ea
%
Notice the nonlocal dependence of $F$ on $G$. The spike in $G$ at
$\phi=0$ (cf.~Fig.~\ref{fig2}a) becomes extremely sharp as
$T{\rightarrow}0$, with little contribution to $F(\phi)$. This,
although easily justified numerically, can be understood by
making an analogy to the case of a superconducting point contact
\cite{new}: $G(\phi)\sim (\Delta/T) e^{-\Delta |\phi|/T}$ near the
spike ($\phi$ measured from the center of the spike), leading to
a contribution to $F(\phi)$ proportional to $\sqrt{T/\Delta}$,
which is negligible as $T{\rightarrow}0$. Numerical integration
shows an even smaller effect.

In the quantum regime, one can truncate the Hilbert space to the
left and right degenerate states, $|L\rangle, |R\rangle$, of the
double-well potential, with $ \langle L|F(\phi ) |R \rangle \simeq
0$ and $\langle R|F(\phi ) |R \rangle \simeq - \langle L|F(\phi )
|L \rangle \simeq F(\phi_0 ).$ Equation~(\ref{H}) then gives the
effective two-state Hamiltonian of the system (at the
degeneracy), coupled to the heat bath, as
\be
 H_{\rm eff} = -{\delta E \over 2} \sigma_x - F(\phi_0) X
 \sigma_z,  \label{Heff}
\ee
where $\delta E$ is the energy splitting between the two lowest
energy states of the system.  The dephasing rate is proportional
to the spectrum of heat bath fluctuations, $S(\omega ) = {\rm Im}
[D(\omega )] \coth(\omega /2T),$ taken at the resonance frequency
of the two-level system [see, e.g., Eq. (3.11) in
Ref.~\onlinecite{leggett}]. Then, the dephasing time
$\tau_\varphi$ due to coupling to the Ohmic heat bath is obtained
from
\ba
 \tau_\varphi^{-1} = \alpha F(\phi_0)^2 \delta E \coth {\delta E \over
 2T}. \label{tau}
\ea
Numerical calculation at $T=10^{-4}T_c$ gives $\int_0^{\pi/2}
\sqrt{G} d\phi \approx 0.38 \sqrt{G_0}$, for $\eta=0.01T_c$.
Substituting into (\ref{F}) and (\ref{tau}) and restoring
$\hbar$, we find $\tau_\varphi^{-1} \approx 0.023 R_Q G_0 \delta E
/\hbar$, where $R_Q{=} h/(2e)^2 {\approx} 6.45$ k$\Omega$ is the
quantum resistance.


For quantitative estimation of $\tau_\varphi$, we need to know the
value of $G_0$. We extract $G_0$ from the Josephson critical
current density whose value is available from experiment
\cite{HM}: $j_c \sim 10^2{-}10^4$~A/cm$^2$. 
Our calculations at $T=0.05 T_c$ (close to 4.2K where most
experiments are performed) show a critical current $I_c \approx
0.08 G_0 T_c/e$, almost independent of $\eta$. For a submicron
junction of area $S\sim 0.01\ \mu$m$^2$, we obtain (taking $T_c
\approx 100$~K) $G_0 \sim 10^{-5}{-}10^{-3}~\mho$. Assuming
$\delta E/h\sim 1$~GHz, we find $\tau_\phi \sim 1{-}100$~ns (qubit
quality factor $Q\sim 1{-}100$). Smaller $\eta$ will result in a
larger $\tau_\phi$. It is therefore desirable to use materials
with low disorder and junctions with small roughness. Depending
on the parameters, it is possible to observe signatures of
quantum behavior (e.g. coherent tunneling), although the
decoherence time may not be long enough for quantum computation.

Realistic junctions, in general, suffer from finite reflectivity,
roughness, and faceting \cite{note3}. The effect of roughness, to
some extent, is similar to that of $\eta$; it broadens the MGS,
increasing their contribution to $G$ as they split
\cite{golubov}. Imperfect transparency also affects the MGS in a
nontrivial way, influencing $G$. Presence of a subdominant order
parameter at the junction, although yet unjustified
experimentally \cite{NV}, can enhance the quasiparticle
resistance by opening a gap at the nodes. A small bulk size can
also have a similar effect by quantizing momentum along the nodal
directions. Most of the above mentioned effects can be studied
within the framework of the present formalism and are the subject
of further investigations.

Other sources, such as fluctuations of the external fields,
coupling of bulk nodal quasiparticles to the electromagnetic
field produced by the qubit (due to e.g. spontaneous currents),
coupling to nuclear spins or paramagnetic impurities, background
charge fluctuations, etc. will also contribute to the decoherence.

It is our pleasure to thank  Ya.~Fominov, A.A. Golubov, W.N.
Hardy, Y. Imry, A. Maassen van den Brink, N. Schopohl,
P.E.C.~Stamp, and A.M. Zagoskin for helpful discussions, A.Ya.
Tzalenchuk for experimental information, and A.J. Leggett for
pointing out the necessity for nonlinear coupling to the heat
bath.


\begin{thebibliography}{99}
\bibitem{eta} Here $\eta$ is a phenomenological
number to be extracted from experiment. It can also be calculated
microscopically, taking into account electron-phonon and impurity
scattering processes; the former makes it $T$-dependent.
\bibitem{note} Notice $V_0$ only appears in the calculation of
$G$; for decoherence calculation, no driving voltage is assumed
(e.g. quantum beating). Even in the presence of a Rabi signal,
since the Rabi frequency $\omega_R$ ($\sim eV_0$) is much smaller
than $\omega$ ($= \delta E$), the condition can be easily met.
\bibitem{xi} In general $\langle \xi_\phi \rangle \ne 0$,
and therefore (\ref{langevin}) is not exactly the classical
Langevin equation. When $T$ is larger than the characteristic
frequencies of the system, it is easy to show that $\langle
\xi_\phi \rangle \approx 0$. At small $T$, however, it stays
finite and adds nonvanishing quantum corrections to
(\ref{langevin}). Study of those are beyond the scope of the
present article.
\bibitem{note3} For small (submicron) junctions,
faceting may only create uncertainty in the orientation of the
boundary.
\end{thebibliography}
\end{document}